\documentclass[aps,prl,twocolumn,nofootinbib,floatfix,10pt]{revtex4-2}

\usepackage{amsmath,mathtools}
\usepackage{amssymb}
\usepackage{wasysym}
\usepackage{graphicx}
\usepackage{color,soul}
\usepackage{physics}
\usepackage{siunitx}
\usepackage{placeins}
\usepackage{dsfont}
\usepackage{float}
\usepackage[english]{babel}
\usepackage{blindtext}
\usepackage[english,nomargin,inline,marginclue,draft]{fixme}
\pdfpageheight\paperheight
\pdfpagewidth\paperwidth

\usepackage[colorlinks,linkcolor=blue,anchorcolor=blue,citecolor=blue,urlcolor=blue]{hyperref}
\usepackage{printlen}
\usepackage{ulem}
            
\fxusetheme{colorsig}
\FXRegisterAuthor{cg}{acg}{CG}  
\FXRegisterAuthor{th}{ath}{\color{blue}TH}  
\FXRegisterAuthor{ib}{aib}{\color{red}IB} 
\FXRegisterAuthor{sh}{ash}{\color{cyan}SH} 
\FXRegisterAuthor{db}{adb}{\color{green}DB} 
\FXRegisterAuthor{ps}{aps}{PS}
\makeatletter
\renewcommand*\FXLayoutInline[3]{%
  {\@fxuseface{inline}\ignorespaces{\color{fx#1}[#3: #2]}}}
\makeatother

\long\def\symbolfootnote[#1]#2{\begingroup%
\def\thefootnote{\fnsymbol{footnote}}\footnotetext[#1]{#2}\endgroup}

\def\nobreakbefore{%
  \relax\ifvmode\else
    \ifhmode
      \ifdim\lastskip > 0pt\relax
        \unskip\nobreakspace
      \else 
        \nobreakspace
      \fi
    \fi
  \fi
}
\let\oldcite\cite
\renewcommand\cite{\nobreakbefore\oldcite}

\newcommand{\eqdef}{\mathrel{\mathop{=}\mkern-4mu\vcentcolon}}

\begin{document}

\title{Time series learning in a many-body Rydberg system with emergent collective amplification}

\author{Zongkai Liu$^{1,2,\star,\S}$}

\author{Qiming Ren$^{1,2,\S}$}

\author{Chris Nill$^{3,4,\dagger,\S}$}

\author{Albert Cabot$^{3,7}$}

\author{Wei Xia$^{5}$}

\author{Yanjie Tong$^{1,2}$}

\author{Huizhen Wang$^{1,2}$}

\author{Wenguang Yang$^{1,2}$}

\author{Junyao Xie$^{1,2}$}

\author{Mingyong Jing$^{1,2}$}

\author{Hao Zhang$^{1,2}$}

\author{Liantuan Xiao$^{1,2}$}

\author{Suotang Jia$^{1,2}$}

\author{Igor Lesanovsky$^{3,6}$}

\author{Linjie Zhang$^{1,2,\ddagger}$}

\affiliation{$^1$State Key Laboratory of Quantum Optics Technologies and Devices, Institute of Laser Spectroscopy, Shanxi University, Taiyuan 030006, China .}

\affiliation{$^2$Collaborative Innovation Center of Extreme Optics, Shanxi University, Taiyuan, Shanxi 030006, China.}

\affiliation{$^3$Institut für Theoretische Physik and Center for Integrated Quantum Science and Technology (IQST), Eberhard Karls Universität Tübingen, Auf der Morgenstelle, 14, 72076, Tübingen, Germany}

\affiliation{$^4$ Institute for Applied Physics, University of Bonn, Wegelerstraße 8, 53115 Bonn, Germany}

\affiliation{$^5$Department of Physics and The Hong Kong Institute of Quantum Information Science and Technology, The Chinese University of Hong Kong, Shatin, New Territories, Hong Kong, 999077, China}

\affiliation{$^6$School of Physics and Astronomy and Centre for the Mathematics and Theoretical Physics of Quantum
Non-Equilibrium Systems, The University of Nottingham, Nottingham, NG7 2RD, United Kingdom}

\affiliation{$^7$Institute for Cross-Disciplinary Physics and Complex Systems (IFISC) UIB-CSIC, Campus Universitat Illes Balears, 07122, Palma de Mallorca, Spain.}

\date{\today}
\symbolfootnote[1]{lzk1997@sxu.edu.cn}
\symbolfootnote[2]{chris.nill@uni-tuebingen.de}
\symbolfootnote[3]{zlj@sxu.edu.cn}
\
\symbolfootnote[4]{Z.L, Q.R, and C.N contribute equally to this work.}

\maketitle
\textbf{Interacting Rydberg atoms constitute a versatile platform for the realization of non-equilibrium states of matter. Close to phase transitions, they respond collectively to external perturbations, which can be harnessed for technological applications in the domain of quantum metrology and sensing.
Owing to the controllable complexity and straightforward interpretability of Rydberg atoms, we can observe and tune the emergent collective amplification.
Here, we investigate the application of an interacting Rydberg vapour for the purpose of time series prediction. The vapour is driven by a laser field whose Rabi frequency is modulated in order to input the time series. 
We find that close to a non-equilibrium phase transition, where collective effects are amplified, the capability of the system to learn the input becomes enhanced.
This is reflected in an increase of the accuracy with which future values of the time series can be predicted.
Using the Lorenz time series and temperature data as examples, our work demonstrates how emergent phenomena enhance the capability of noisy many-body systems for data processing and forecasting.}

\begin{figure*}[ht]
\includegraphics[width=\linewidth]{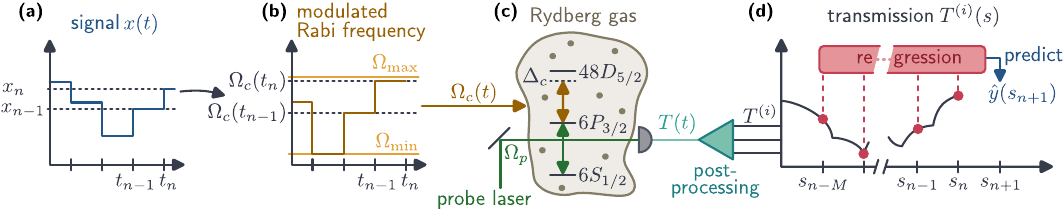}
    \caption{\textbf{Time series prediction with a Rydberg vapour.}
    (a)The time-dependent input signal $x(t)$ is sampled at discrete time steps $t_n$, yielding a step-wise constant function with values $x_n=x(t_n)$.
    (b)The signal is converted into the Rabi frequency $\Omega_c(t_n)$ of the coupling laser, which couples the intermediate state $6 P_{3/2}$ to the Rydberg state $48 D_{5/2}$ detuned by $\Delta_c$. This Rabi frequency ranges within the interval $[\Omega_\mathrm{min}, \Omega_\mathrm{max}]$.
    (c)The caesium~(Cs) atoms are first excited by a probe laser with a constant Rabi frequency $\Omega_p$, which drives the transition from the ground state $6 S_{1/2}$ to the intermediate $6 P_{3/2}$. Subsequently, the atoms are excited by the modulated coupling laser.
    The optical response of the Cs vapour is measured by monitoring the transmission $T(t)$ of this probe laser.
    (d)The measured transmission $T(t)$ is post-processed after measurement using a Savitzky-Golay filter and down-sampled by a factor of $20$, yielding the transmission signals $T^{(i)}=T(t_{i+20n})$. Here, $i\in[1,20]$ labels the index of the downsampled time series and $s_n=t_{i+20n}$ denotes the corresponding time.
    The transmission signal within the time interval $[s_{n-M}, s_n]$ is used to predict the signal $x(s_{n+1})$ (denoted as $\hat y(s_{n+1})$).
    This prediction is generated by a linear regression layer, which was previously trained (see text for details).
    }
    \label{fig:setup}
\end{figure*}

Computation, information processing, and dynamics in many-body systems are inherently intertwined. One paradigmatic example is the Hopfield network~\cite{Krotov2023Hopfield}, whose architecture, energy function, and memory retrieval are directly derived from the Ising model. Other examples include swarm intelligence algorithms, such as particle swarm optimization~\cite{Kennedy.1995Particle,Parsopoulos2002PSO,wang2018particle} and ant colony optimization~\cite{Dorigo2006ACO,DORIGO2005243}. These systems exploit collective interactions across scales to exhibit emergent computational capabilities, where nonlinear coupling between microscopic degrees of freedom generates information-processing capabilities. This interplay underscores both the potential of physical systems as computational platforms and the need to decode how collective behaviour enables specific computational tasks.

A particularly relevant and challenging task is time-series prediction of complex systems, for example chaotic dynamics such as the Lorenz attractor~\cite{Ghys2013, Shen2023} or weather forecasting~\cite{Mammedov2022}.
A computational framework that has proven effective for such problems is reservoir computing~\cite{Jaeger2001,tanaka2019recent,Nakajima_2020, Wringe2025}.
In reservoir computing, only a linear readout layer is trained, while the high-dimensional, nonlinear reservoir dynamics remain fixed. This significantly simplifies training, as it avoids tuning a large number of internal degrees of freedom and instead leverages the computational power of the intrinsic dynamics of the system, provided they are sufficiently rich~\cite{Dale2019,Carroll2022,gauthier2021next}.
Reservoir computing has been implemented in a variety of settings, such as in echo state networks~\cite{Ozturk2007AnalysisESNs,Sun2024SystematicReview,Herbert2004ESMs} and liquid state machines ~\cite{LSMs2011Wolfgang} as well as in physical systems such as delay-based electronic circuit reservoirs~\cite{Soriano2015DBRC}, and quantum reservoir computing platforms~\cite{Fujii2017,Mujal2021a,Sannia2024}.
Despite this diversity, understanding the fundamental mechanisms that endow complex physical reservoirs with learning and prediction capabilities remains subject to current research~\cite{Yan2024}.

Thermal Rydberg vapours are a promising platform in this context. These many-body systems feature strong interatomic interactions and coexisting drive-dissipation processes, whose competition gives rise to a rich phase diagram \cite{ates2006strong,marcuzzi2014universal,lesanovsky2014structures,ding2019Phase,klocke2021hydrodynamic} and a wealth of emergent non-equilibrium phenomena, including collective jumps\cite{carr2013nonequilibrium,PhysRevResearch.6.L032069}, ergodicity breaking\cite{ding2023ergodicity,wadenpfuhl2023emergence}, self-induced
transparency\cite{bai2020selfinduced}, and dissipative time crystals\cite{Wu2024Dissipative}. These features make Rydberg ensembles attractive candidates as platforms for reservoir computing, and, at the same time, ideal systems for probing the microscopic origin of learning capabilities in complex many-body dynamics at the atomic level.

In this work, we exploit the response of a Rydberg vapour as a physical reservoir for time-series prediction.
The experimental control over the collective behaviour of the system enables us to systematically analyse how its internal dynamics give rise to learning capabilities.
We find that the learning performance is improved in the vicinity of a phase transition, where the collective response is strongly amplified in the Rydberg vapour.
This behaviour is in good agreement with a corresponding theoretical minimal model, providing insight into how critical many-body dynamics augment the computational power of physical reservoirs.

\section{Results}
\begin{figure*}[t]
    \includegraphics[width=1\linewidth]{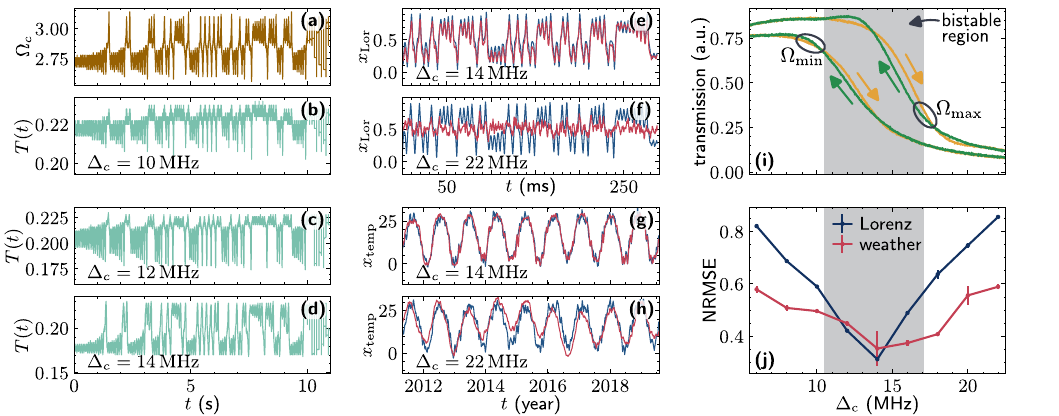}\caption{\textbf{Time series prediction.}
    \textbf{(a)}~Amplitude modulated signal $\Omega_c/2\pi$ in MHz which is input into the Rydberg system. 
    \textbf{(b-d)}~Output transmission signal $T(t)$ in arbitrary units recorded for different detunings $\Delta_c$.
    The detuning is chosen such that the system is out of (panel b, its $\Delta_c$ out of the hysteresis region in panel i) and within the bistable region (panels c-d, $\Delta_c$ inside the hysteresis region in panel i).
    \textbf{(e, f)}~Time series prediction for the Lorenz series $x_\mathrm{Lor}$. A Rydberg vapour operating in the bistable region (e) results in a better prediction (red curve), compared to outside (f). The ground truth is shown in blue.
    \textbf{(g-h)}~Prediction results on daily temperature of Beijing for detuning $\Delta_c$, near (g) and out (h) of the bistable region.
    \textbf{(i)}~The lower yellow and lower green curves correspond to the minimal Rabi frequency $\Omega_{\mathrm{min}}$ of the coupling laser but with different scan directions of the detuning $\Delta_c$ (as indicated by arrows and colours). The upper yellow and upper green curves correspond to $\Omega_{\mathrm{max}}$. 
    A closed hysteresis loop formed by the two scan directions characterizes the bistable phenomenon.
    The detuning interval where the bistability occurs is marked by a gray shaded box.
    \textbf{(j)}~Normalised root mean square error (NRMSE) for varying detuning $\Delta_c$ on Lorenz and weather temperature time series prediction tasks. Inside the bistable region (gray shaded box), the NRMSE is lower than outside.
    The plotted NRMSE values correspond to the average over the 20 sampled time series, while the error bar corresponds to their standard deviation (see text for details).
    }
    \label{fig:prediction}
\end{figure*}

Before discussing the detailed experimental results, we briefly outline the overall aim and approach of our study. 
The experiment is designed to test whether the vapour of interacting Rydberg atoms can process and predict time-dependent input signals by exploiting its intrinsic nonlinear and collective dynamics, see Fig.~\ref{fig:setup}. 
To achieve this, we encode time-series data $x(t)$ in an amplitude modulation of the coupling laser’s Rabi frequency $\Omega_c$.
The coupling laser drives the atomic ensemble under fixed detuning $\Delta_c$ and serves as input signal.
The transmission of a probe laser serves as the measurable output that reflects how the Rydberg vapour responds and processes the input information. 
We record this transmission signal as a time series and use it as input to a simple linear regression model that is trained to predict future values of the original time series $x(t)$. 
By comparing the system’s behaviour for different detunings $\Delta_c$, particularly within and outside regions of bistability, we can assess how collective nonlinear effects of the Rydberg vapour influence its predictive capability.

\textbf{Experimental setup.}
In our experiment, caesium~(Cs) atoms are placed inside a $(5 \times 5 \times 5)\,\si{\centi\meter\cubed}$ heated cube glass cell,
with the temperature maintained at \SI{47.0}{\celsius}, corresponding to an atomic density of \SI{3.61e11}{\per\cubic\centi\meter}~\cite{vsibalic2017arc}.
As shown in Fig.~\ref{fig:setup}, a probe laser with wavelength \SI{852}{\nano\meter} and Rabi frequency $\Omega_p/2\pi \sim \SI{84.92}{\mega\hertz}$ drives the $\ket{6S_{1/2}, F=4} \leftrightarrow \ket{6P_{3/2}, F=5}$ transition (D2 line).
The coupling laser with wavelength \SI{510}{\nano\meter} and Rabi frequency $\Omega_c$ is used to excite atoms to the $\ket{48D_{5/2}}$ Rydberg state.
The coupling and probe lasers are aligned counter propagating in the atomic vapour cell to partially suppress Doppler broadening, forming an electromagnetically induced transparency (EIT) configuration.
In this configuration, the transmission of the probe laser through the Rydberg vapour is controlled by the coupling laser~\cite{Fleischhauer2005eit}.
The Rabi frequency $\Omega_c$ of the coupling laser is amplitude-modulated using an acousto-optic modulator~(AOM) and an arbitrary waveform generator~(AWG), which operates between minimal $\Omega_\mathrm{min}/2\pi\sim \SI{2.56}{\mega\hertz}$ and peak Rabi frequency $\Omega_\mathrm{max}/2\pi \sim \SI{3.14}{\mega\hertz}$. The Cs D2 line is utilized to lock the probe laser frequency, while the coupling laser frequency is stabilized to a super-stable cavity with finesse \num{2e5}.
The frequency difference between coupling (probe) light and the corresponding transition $\ket{6P_{3/2}}\leftrightarrow$$\ket{48D_{5/2}}$ ($\ket{6\mathrm{S}_{1/2}} \leftrightarrow \ket{6P_{3/2}}$) is defined as the detuning $\Delta_c$ ($\Delta_p$). To measure the EIT spectrum, the frequency of the coupling light is scanned with $\Delta_c$ swept back and forth from $\SIrange{0}{20}{\mega\hertz}$, while the probe light is kept resonant with the D2 line ($\Delta_p=0$), and the probe transmission is recorded by a photodetector.

\textbf{Time series prediction.}
We benchmark the predictive capability of the Rydberg ensemble using two qualitatively different time series: the $x$-component of the Lorenz attractor, $x_{\mathrm{Lor}}$ (see Methods), and the daily maximum temperature in Beijing date from 2011-01-01 to 2020-03-23, $x_{\mathrm{temp}}$~\cite{Zenodo}.
These examples probe the system's performance and generalization capability across different types of data, specifically chaotic and stochastic signals.

The input signal $x(t)$ is first discretized at equidistant times $t_n$ to obtain $x_n = x(t_n)$.
Each value $x_n$ is then mapped linearly to the interval $[\Omega_\mathrm{min}, \Omega_\mathrm{max}]$ and encoded as a time amplitude modulation in the coupling Rabi frequency $\Omega_c(t_n) \in [\Omega_\mathrm{min}, \Omega_\mathrm{max}]$, with modulation rate of \SI{4}{\kilo\hertz}, see Fig.~\ref{fig:setup}.
For the Lorenz attractor, we show a representative input signal $\Omega_c(t)$ in Fig.~\ref{fig:prediction} (a).
The modulated coupling laser drives the Rydberg vapour at a fixed detuning $\Delta_c$, while the transmission $T(t)$ of the probe laser is recorded continuously. 
Sampling at times $t_n$ yields a discrete output sequence $T(t_n)$ that reflects the system's response to the input.
Representative transmission traces for three characteristic detunings $\Delta_c = 10, 12$, and $\SI{14}{\mega\hertz}$ are shown in Figs.~\ref{fig:prediction}(b--d), respectively: outside the bistable regime [panel (b)] and within the bistable region [panels (c,d)].
Specifically, the bistable region is defined by the closed hysteresis interval formed by the two scan directions of the coupling laser (shaded interval in panel (i)).
When $\Delta_c$ lies within this shaded interval, the system operates within the bistable region (e.g., $\Delta_c=12,\SI{14}{\mega\hertz}$), whereas detunings outside this interval (e.g., $\Delta_c=\SI{10}{\mega\hertz}$) correspond to the non-bistable regime.
Near bistability, the peak-to-peak amplitude of the transmission is visibly enhanced, consistent with the increased collective gain inside the bistable region.

For the subsequent analysis, the raw transmission data $T(t_n)$ are first processed with a Savitzky--Golay filter (window size 10, polynomial order 3~\cite{CHEN2004332,CAO2018244}) to suppress high-frequency noise.
For a discussion of causality of this filtering process we refer to Methods.
To obtain several time series from a single run, we generate 20 downsampled transmission series by taking every 20th point with different initial offsets, see Fig.~\ref{fig:setup}.
Thus, each sub-series $T^{(i)}$ (with $i = 1,\dots,20$) contains one out of every 20 points of the original record and has one twentieth of its length.
For example, for the sub series $T^{(1)}$, the downsampled series contains 1st, 21th, 41th,..., ($20t_n+i$)-th points of the original record.

The prediction task is formulated using a sliding window over each downsampled series.
We fix a window length $M = 200$ and, for each series $i$, use $M$ consecutive transmission values as input and the next value of the downsampled time series as the prediction target.
Writing
\begin{equation}
  \vec{T}^{(i)}_k = \bigl(T^{(i)}_k, T^{(i)}_{k+1}, \dots, T^{(i)}_{k+M-1} \bigr)^{\mathsf{T}}
\end{equation}
for the $k$-th input window, the corresponding target is chosen as
\begin{equation}
  y^{(i)}_k = x\bigl(t_{i+20(k+M)}\bigr).
\end{equation}
Shifting the window generates the full set of input–target pairs $\bigl(\vec{T}^{(i)}_k, y^{(i)}_k\bigr)$ which are indexed by $k$.
Thus, the input-target pairs include transmission data of the probe laser in $\vec{T}^{(i)}_k$ as well as the consecutive point of the time series $y^{(i)}_k$ which is proportional to the Rabi frequency of the coupling laser $\Omega_c$.

We use 70\% of the input–target pairs for training and the remaining 30\% for testing.
For each subseries $i$, we train a linear regression model that maps the $M$-dimensional transmission window $\vec{T}^{(i)}_k$ to a scalar prediction $\hat y^{(i)}_m$ of the next value of the original time series.
Introducing a weight vector $\vec{w}^{(i)}$ and a scalar bias $b^{(i)}$, the linear regression model is defined as
$\hat y^{(i)}_m = \vec{T}^{(i)}_m \cdot \vec{w}^{(i)} + b^{(i)}$.
The parameters $\vec{w}^{(i)}$ and $b^{(i)}$ are obtained by minimizing the mean-squared error (MSE) on the training set,
$\mathrm{MSE}^{(i)}_\mathrm{train} = \frac{1}{N^{(i)}_\mathrm{train}} \sum_{k \in \mathrm{train}} \bigl(y^{(i)}_k - \hat y^{(i)}_k\bigr)^2$,
where $N^{(i)}_\mathrm{train}$ denotes the number of training samples in series $i$.

It is noted that no nonlinear activation functions are used in the readout layer. 
The readout operations are limited to linear regression (least squares method) and matrix multiplication, which are strictly linear calculations.
Therefore, all the nonlinear processing of the input signal observed in our experiments originates from the intrinsic dynamics of the Rydberg vapor, which is characterized by the hysteresis loop in the system's response (Fig. 2i).

Figures~\ref{fig:prediction}(e--h) show representative prediction results.
When the system is operated within the bistable regime [Figs.~\ref{fig:prediction}(e), (g)], the predicted trajectories closely follow the Lorenz signal and the temperature record, including their short-time fluctuations.
Outside the bistable region [Figs.~\ref{fig:prediction}(f), (h)], the predictions deviate substantially from the ground truth, indicating a reduced learning performance.
\begin{figure}
    \centering
    \includegraphics[width=\linewidth]{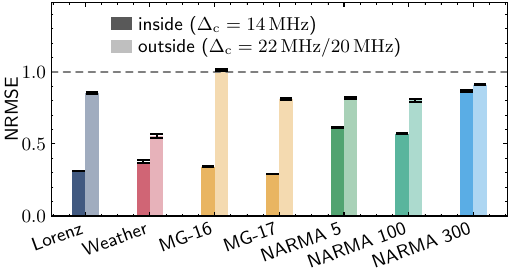}
    \caption{\textbf{Prediction performance inside and outside the bistable region.} Normalised root mean square error (NRMSE) for all benchmark tasks, comparing operation inside the bistable region ($\Delta_\mathrm{c} = 14\,$MHz) and outside ($\Delta_\mathrm{c} = 22\,$MHz, or $20\,$MHz for NARMA). For all tasks --- Lorenz, weather temperature, Mackey--Glass ($\tau_\mathrm{MG} = 16, 17$), and NARMA-$n$ ($n = 5, 100, 300$) --- the prediction error is lower inside the bistable region, where the collective susceptibility of the Rydberg vapour is enhanced. The NRMSE values correspond to the average over 20 downsampled sub-series; error bars denote their standard deviation.
    }
    \label{fig:benchmarks}
\end{figure}

The dependence on detuning is summarized in Figs.~\ref{fig:prediction}(i,j).
Figure~\ref{fig:prediction}(i) displays the EIT spectrum as a function of the coupling detuning $\Delta_c$ for minimal ($\Omega_\mathrm{min}$) and maximal ($\Omega_\mathrm{max}$) Rabi frequency.
For each value of $\Omega_c$, we scan $\Delta_c$ once from low to high frequency (yellow curves) and once in the opposite direction (green), and record the probe transmission as output.
The upper pair corresponds to $\Omega_\mathrm{max}$, and the lower pair to $\Omega_\mathrm{min}$.
Since the modulation range of $\Omega_c$ is $\Omega_c \in [\Omega_\mathrm{min}, \Omega_\mathrm{max}]$, a transmission region exists during the modulation of the Rabi frequency $\Omega_c$; for each detuning $\Delta_c$, this region is bounded by the upper pair and lower pairs.
These two branches of the transmission curve form a closed hysteresis loop: in the central gray-shaded interval, the measured transmission at a given $\Delta_c$ depends on the scan history, which signals the presence of two coexisting stable states of the Rydberg vapour.
Consequently, the detuning $\Delta_c$ serves as a control parameter that characterizes the proximity of the system to the bistable region and adjusts the strength of the nonlinearity.

This bistable region is located precisely in the parameter window in which the collective gain is largest and where the prediction performance of the system is found to be optimal.
This is quantified by the normalised root mean squared error~(NRMSE)~\cite{Wringe2025}, defined as
\begin{equation}
 \operatorname{NRMSE}= \mathrm{avg}_i\left(\sqrt{ \frac{\sum_{k=1}^{N^{(i)}_\mathrm{pred}} \left( {y}^{(i)}_k - \hat{y}^{(i)}_k \right)^2}{\sum_{k=1}^{N^{(i)}_\mathrm{pred}} \left( {y}^{(i)}_k - \bar{y}^{(i)} \right)^2} }\right),
 \label{eq:nrmse}
 \end{equation}
 where  $\bar{y}^{(i)}$ is the temporal mean of the target sequence.

Figure~\ref{fig:prediction}(j) presents the NRMSE as a function of detuning: the error exhibits a clear minimum inside the bistable window and increases again outside it.
The error bars denote the standard deviation over the twenty downsampled sub-series.

In summary, the Rydberg vapour shows its best prediction performance on both tasks when operated near the bistable regime adjusted by detuning $\Delta_c$, where the collective gain is largest.
The correlation between the EIT hysteresis and the minimum NRMSE points to emergent collective effects, rather than single-atom physics, as the key resource for enhanced learning capacity.

\textbf{Standard benchmark tasks.}
To quantify the nonlinear capacity of the Rydberg reservoir beyond the Lorenz and temperature tasks, we evaluate its performance on multiple standard benchmarks from the reservoir computing literature: the Mackey-Glass~(MG) chaotic time series and the NARMA-$n$ task, see Fig.~\ref{fig:benchmarks}.

The Mackey-Glass system is a delay-differential equation that generates chaotic dynamics for sufficiently large delay parameters (see Methods). We tested our system with delays $\tau_\mathrm{MG} = 16$ and $17$. Operating inside the bistable region, the reservoir successfully tracks the chaotic dynamics, achieving NRMSE values of $0.34$ and $0.29$, respectively, which is a substantial improvement over the values obtained outside the bistable region ($\Delta_c = 22\,$MHz), where NRMSE values are $1.01$ and $0.81$, respectively.

The NARMA-$n$ (nonlinear auto-regressive moving average) benchmark is the established standard for evaluating a reservoir's ability to nonlinearly transform past inputs (see Methods).
In Fig.~\ref{fig:benchmarks} we evaluated the complexities $n = 5$, $100$, and $300$ at two detunings, one within and one outside the bistable region. Further data for additional $n$ and detuning values are given in the Methods.
Inside the bistable region, where the collective susceptibility is enhanced, the system maintains a low prediction error (NRMSE $\approx 0.61$ for NARMA-5 and NRMSE $\approx 0.57$ for NARMA-100). Outside the bistable region ($\Delta_\mathrm{c} = 20\,$MHz), the performance is degraded (NRMSE $\approx 0.82$ for NARMA-5 and NRMSE $\approx 0.80$ for NARMA-100). For NARMA-300, which requires memory exceeding the $M=200$ readout window boundary, the performance collapses to NRMSE $\approx 0.87$ (inside) and NRMSE $\approx 0.91$ (outside), demonstrating that the system memory is window-bounded (see Methods). These results show that the system operates in a high-gain, noise-resilient regime where the enhanced collective susceptibility provides the main performance boost.

\begin{figure}
    \centering
    \includegraphics[width=0.95\linewidth]{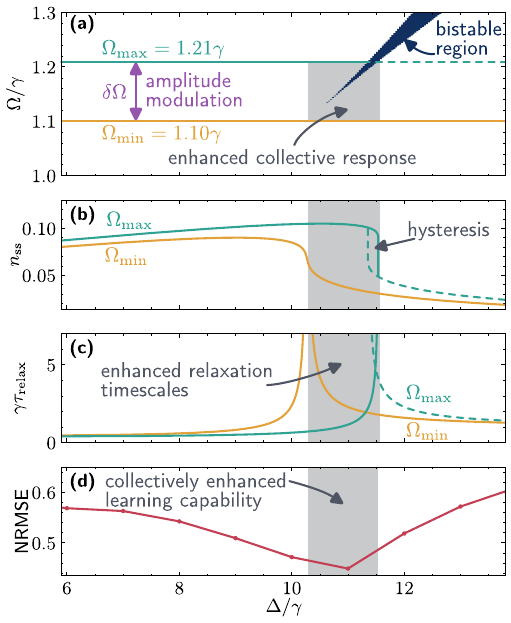}
    \caption{
    \textbf{Mean-field dynamical response and learning capacity.}
    \textbf{(a)}~Phase diagram of the mean-field equation~(\ref{eq:n-dot-eliminated}). The blue region marks the bistable domain where two stable stationary solutions exist. The yellow (green) horizontal lines indicate the minimum (maximum) Rabi frequency of the input signal of the learning protocol, see panel (d). The gray shadowed area indicates the region in which collective effects are most prominent for the considered Rabi frequencies. Its left (right) boundary corresponds to the detuning at which the relaxation time for the minimum (maximum) Rabi frequency displays a maximum, see panel (c). 
    \textbf{(b)}~Stationary Rydberg density $n_\mathrm{ss}$ for varying detuning along the horizontal cuts $\Omega/\gamma=1.1$ and $\Omega/\gamma=1.21$. For the case $\Omega/\gamma=1.21$ hysteresis is observed. We denote the solution with the lower density with a dashed line.
    \textbf{(c)}~Relaxation time \( \tau_\mathrm{relax} \) along the same cuts as in (b) for the stable stationary solutions. Within the bistable region, two relaxation times are possible, one for each stable solution. The solid (dashed) line corresponds to the solution with higher (lower) density. The longest relaxation times occur near the boundary of the bistable region. 
    \textbf{(d)}~Normalised root mean square error~(NRMSE) for the prediction of the Lorenz input signal fixing the parameters to $\Omega/\gamma=1.1$, modulation amplitude $\delta \Omega/\Omega=0.1$, modulation period $\gamma T=20$, noise strength $D/\gamma=0.0001$ and for different detunings (points). In all panels $V/\gamma=100$ and $\gamma_\mathrm{d}/\gamma=10$.
    }
    \label{fig:meanfield-and-response}
\end{figure}
\textbf{Theoretical model.}
The experimental results we have just presented, reveal a correlation between improved learning capability and the regime in which the Rydberg vapour displays bistability.
In order to gain understanding on this relation, we are going to describe the Rydberg vapour using a mean-field model, see e.g. Refs.~\cite{carr2013nonequilibrium,marcuzzi2014universal,PhysRevResearch.6.L032069}, and analyse the learning capability of the modelled system.
This minimal theoretical model is designed to capture the essential features of the many-body phase transition and bistable regime relevant to the reservoir computing response. It employs an all-to-all mean-field approximation, a standard description for collective dynamics in Rydberg ensembles, and is derived from the Lindblad quantum master equation. The model describes collective phenomena involving atomic interactions and driven-dissipative processes, without invoking strong quantum correlations or entanglement as the origin of the observed dynamics.

The dilute vapour of laser-driven Rydberg atoms is modelled as \(N\) identical two-level systems with ground \(|g\rangle\) and Rydberg \(|e\rangle\) states (see Methods). The atoms are driven with Rabi frequency $\Omega$ and detuning $\Delta$, while dissipation enters through spontaneous emission with rate $\gamma$ and dephasing with rate $\gamma_\mathrm{d}$. The interaction between the atoms in the Rydberg state is modelled as an all-to-all energy shift term in the Hamiltonian, i.e. $H_\mathrm{int}=\frac{V}{N-1}\sum_{j< k}\dyad{e}{e}_j \dyad{e}{e}_k$. In the thermodynamic limit, the dynamics is described by the mean-field equations of motion for the average population $n(t)=\frac{1}{N}\sum_{j=1}^N\ev*{\dyad{e}_j}$ and optical coherence $q(t)=\frac{1}{N}\sum_{j=1}^N\ev*{\dyad{e}{g}_j}$. Since in a dilute vapour at room temperature $\gamma_\mathrm{d}$ typically exceeds $\gamma$ and $\Omega$, the coherence relaxes on a much shorter timescale $(\dot q\gg \dot n)$, justifying its adiabatic elimination (see Methods).
Therefore, the dynamics is well described by an equation for the average Rydberg population alone. This reads
\begin{equation}
  \dot n =
  -\frac{\Omega^{2}\Gamma\left(n-\frac{1}{2}\right)}
        {\Gamma^{2}+(\Delta-nV)^{2}}
  -\gamma n \eqdef F(n),
  \label{eq:n-dot-eliminated}
 \end{equation}
where we have defined the combined dissipation rate $\Gamma=\tfrac12(\gamma+\gamma_{\mathrm d})$. This nonlinear differential equation captures the interplay of coherent driving, interactions, dephasing, and spontaneous emission. Depending on the parameter values, the model displays one or two stable solutions for the steady-state Rydberg density $n_\mathrm{ss}$, which are shown in Fig.~\ref{fig:meanfield-and-response}(a). 

In the experiment, the input signal is encoded by modulating the Rabi frequency.
The system operates in a regime characterized by the EIT spectrum curves shown in Fig.~\ref{fig:prediction}(i), in which a more pronounced hysteresis is observed for larger Rabi frequencies. We find that the mean-field model displays a similar response in the stationary Rydberg density $n_\mathrm{ss}$ when operating at Rabi frequencies around the lower critical point at which the bistable region emerges. This is illustrated in Fig.~\ref{fig:meanfield-and-response}(b), in which we show the stationary Rydberg densities along the maximum and minimum Rabi frequencies used in the learning protocol. In this region, collective effects result in a strong nonlinear response near the boundaries of the bistable regime, as well as hysteresis.
This strong nonlinearity enhances the system's gain within the shaded region of Fig.~\ref{fig:meanfield-and-response}(b), which, in the presence of thermal fluctuations and noise, improves the effective signal quality.
In the experiment, this enhancement is reflected by the increased peak-to-peak value from Fig.~\ref{fig:prediction}(b) to Fig.~\ref{fig:prediction}(d) (see also Methods).

In order to theoretically analyse the correlation between these collective effects and the enhanced learning capability, we need to complement the mean-field equation~\eqref{eq:n-dot-eliminated} with a stochastic term that models the presence of noise in the experiment. We then consider the following stochastic equation for the Rydberg density: $\dot{n}=F(n)+\sqrt{n D}\xi(t)$, where $\xi(t)$ is a white noise term, and $D$ a constant controlling its overall strength. The input signals to be predicted are then introduced analogously as in the experiment, by modulating the Rabi frequency in a piecewise fashion, with maximum modulation amplitude $\delta \Omega:=\Omega_\mathrm{max}-\Omega_\mathrm{min}=0.1\Omega$ and modulation period $\gamma T =20$, see Fig.~\ref{fig:setup}. The corresponding NRMSE for the Lorenz input is shown in Fig.~\ref{fig:meanfield-and-response}(d) for different $\Delta/\gamma$ along a cut at $\Omega/\gamma=1.1$. The NRMSE displays a dip around $\Delta/\gamma\sim 11$, coinciding with the region in which collective effects are most prominent, see the corresponding shadowed region in Fig.~\ref{fig:meanfield-and-response}(b) and (d), hence supporting our previous hypothesis based on the experimental results, see Fig.~\ref{fig:prediction}(i) and (j).
Therefore, our theoretical findings align with the experimental finding that the learning capability is improved in the bistable region, where collective effects are more prominent.

Experiment and theory also align in the observation that relaxation dynamics slows down in the regime in which the learning capability is improved. Theoretically, characteristic relaxation times, $\tau_\mathrm{relax}$, can be obtained by linearizing Eq.~(\ref{eq:n-dot-eliminated}) around each of its stationary solutions, $n_\mathrm{ss}$ (see Methods). The typical behaviour of the relaxation time within the regime of operation is shown in Fig.~\ref{fig:meanfield-and-response} (c). We observe that near the boundaries of the bistable regime, the so-called spinodal lines \cite{marcuzzi2014universal}, the relaxation times increase considerably, in agreement with experimental observations (see Methods). This prolonged relaxation time is characteristic of complex systems undergoing critical behaviour and is also referred to critical slowing down~\cite{Dakos2008Slowing, Scheffer2009Early, carr2013nonequilibrium, Brookes2021Critical, Zhang2024Early, Wang2025delay}.

\section{Discussion}
In this work, we have experimentally demonstrated that a driven-dissipative Rydberg vapour displays improved time-series prediction performance when operated near a bistable phase transition. This improvement is consistently observed across all benchmark tasks and is reproduced by a mean-field model, establishing a direct link between emergent collective phenomena and learning capability in Rydberg atomic systems.

Our analysis identifies the underlying mechanism: near bistability, the collective susceptibility peaks, providing a large effective gain that amplifies the contrast between different input levels in the output. The signal-to-noise ratio, by contrast, remains approximately flat across detunings (Fig.~\ref{fig:M4_transfer_curves}c) and is not the driver of the improvement. The NARMA benchmarks and IPC analysis further reveal that outside the bistable region the operating point lies on the quadratic tail of the EIT transfer curve, encoding the input through a basis that the linear readout cannot invert; inside the bistable region, the large gain maps inputs onto a nearly linear basis, to which the linear readout is naturally matched.

The enhanced relaxation times observed near the phase boundary (Fig.~\ref{fig:pulse}) are consistent with critical slowing down. However, these timescales ($\sim 0.14\,$ms) remain shorter than the effective sample interval after downsampling ($5\,$ms), so the vapour itself resets between successive input steps and does not provide fading memory through its physical relaxation dynamics. The temporal memory available to the readout layer originates instead from the sliding window of size $M$. The dominant computational resource is therefore the instantaneous nonlinear transformation provided by the collective susceptibility, not dynamical memory of the reservoir.

Although the overall prediction accuracy is not yet competitive with established reservoir computing platforms, which report significantly lower normalised root mean squared errors\cite{Wringe2025, Fujii2017,Mujal2021a,Sannia2024}, our results show that the collectively enhanced gain near a many-body phase transition is a tunable and experimentally accessible computational resource. We stress that this gain mechanism is general rather than specific to Rydberg systems~\cite{Sugano2020,Mujal2021a,Sannia2024}; its distinctive realisation here is as a many-body phase transition that is continuously tunable through a single control parameter.
Future work may implement advanced methods from the reservoir computing toolbox, such as time-multiplexing and feedback ~\cite{Appeltant2011,Zhang2014,Sugano2020}, combined with reduced downsampling, to further improve the prediction performance in this Rydberg vapour setup.
Future research could also investigate the relation between prolonged relaxation time and learning capacity, and the specific mechanistic role of relaxation time in mediating the learning process.

\section{Methods}
\label{sec:methods}

\textbf{Time series generation.}
The Lorenz oscillator $x(t)$ that we used is defined by the differential equation
\begin{align}
    \dot x(t)&= \sigma\,[y(t) - x(t)], \nonumber \\
    \dot y(t)&= \rho\, x(t) - y(t) - x(t) z(t),\\
    \dot z(t)&= x(t)\, y(t) - \beta\, z(t)\nonumber ,
\end{align}
where $\sigma=10$, $\rho=28$, $\beta=8/3$, time step $t_n-t_{n-1}=$ \SI{0.04}{\second}, total time \SI{640}{\second}.
Thus, the Rabi frequency of the coupling laser is modulated by $x(t)$ at a rate of $\SI{4}{\kilo\hertz}$ per bit.
The Mackey-Glass series (MG) is defined by the delay-differential equation
\begin{equation}
    \dot{x}(t) = \frac{a\, x(t-\tau_\mathrm{MG})}{1 + x(t-\tau_\mathrm{MG})^{10}} - b\, x(t),
\end{equation}
with parameters $a = 0.2$, $b = 0.1$, and delay $\tau_\mathrm{MG}$. For $\tau_\mathrm{MG} > 17$ the dynamics becomes chaotic. We tested delays $\tau_\mathrm{MG} = 16$ and $17$.
The NARMA-$n$ (nonlinear auto-regressive moving average) benchmark~\cite{Fujii2017Harnessing} is defined by the recurrence relation
\begin{equation}
    y_{k+1} = \alpha\, y_k + \beta\, y_k \sum_{j=0}^{n-1} y_{k-j} + \gamma\, u_{k-n+1}\, u_k + \delta,
\end{equation}
where $u_k \in [0, 0.5]$ is the input drawn uniformly at random, $\alpha = 0.2$, $\beta = 0.04$, $\gamma = 1.5$, and $\delta = 0.001$. We evaluated complexities $n = 5$, $10$, and $15$.

\textbf{Experimental protocol for benchmark tasks.}
For each benchmark time series (MG and NARMA-$n$), the dataset consists of 3500 training points and 1500 testing points.
To obtain statistically meaningful error bars, we perform 20 realisations, leading to a total dataset size of $10^5$.
Except for the modulation waveform and small shifts of the detuning, all experimental parameters are kept identical to those described in the main text.
 
Before applying the time-dependent modulation, the operating point is set by tuning the detuning $\Delta_c$ in the absence of modulation to maximise the signal-to-noise ratio of the transmission signal.
The modulation sequence is then applied and the corresponding optical transmission is recorded.
After each measurement, the modulation sequence is switched to the next benchmark.
Small drifts in the experimental detuning between runs mean that the system remains within the bistable region but samples slightly different operating points across realisations, contributing to the spread captured by the error bars in Fig.~\ref{fig:benchmarks}.

\textbf{Data processing and causality.}
Prior to model training, the raw transmission data $T(t_n)$ are smoothed with a symmetric Savitzky--Golay filter (window size 10, polynomial order 3~\cite{CHEN2004332,CAO2018244}) applied offline to the full recorded trace.
Because this filter is symmetric (non-causal), we verified that it does not bias the prediction results by conducting a control experiment using a strictly causal finite impulse response (FIR) filter.
The prediction performance was highly consistent between the two filters (with a slight reduction in NRMSE for the causal filter), confirming that the offline symmetric smoothing has a negligible impact on the results.
 
All time-series prediction uses the causal sliding-window strategy described in the main text: only transmission values at times $s_n$ and earlier are used to predict the input at time $s_{n+1}$.

\begin{figure*}[t]
    \includegraphics[width=\linewidth]{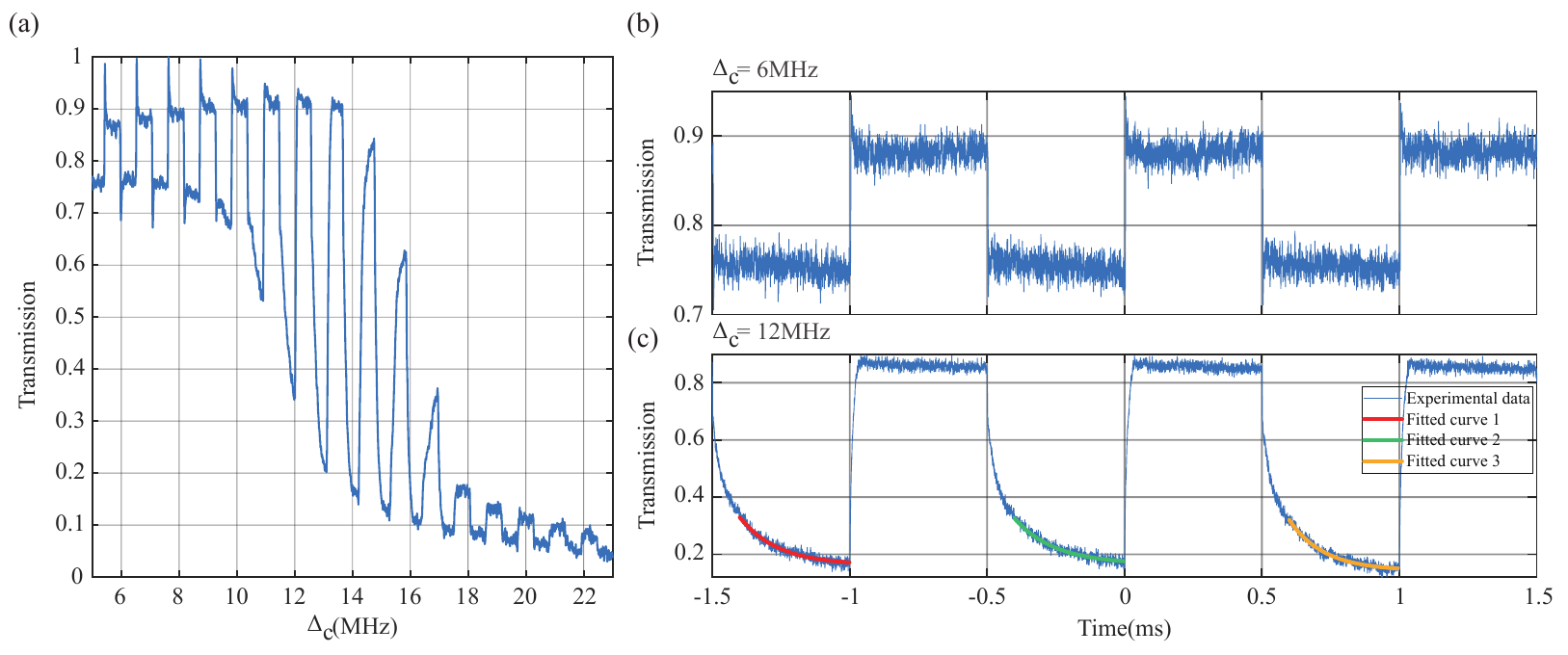}
    \caption{\textbf{Experimental analysis of the relaxation times.} Measurement of the transmission of the probe laser when the Rabi frequency $\Omega_c$ of the coupling light is amplitude modulated by a square pulse with frequency $\SI{1}{\kilo\hertz}$, duty cycle 50\%, and modification depth 15\%. In panel (a), the detuning of the coupling light $\Delta_c$ is swept from $\SIrange{6}{22}{\mega\hertz}$ linearly in time. Instead, in panels (b) and (c) the detuning is fixed to $\SI{6}{\mega\hertz}$, and $\SI{12}{\mega\hertz}$, respectively. Panel (b) corresponds to a point outside the bistable regime, while panel (c) to a point within the bistable regime. In panel (c) an exponential relaxation is fitted to the experimental data (see text) at three different periods obtaining the relaxation times $\tau_1=0.139\,\si{ms}$, $\tau_2=0.148\,\si{ms}$ and $\tau_3=0.125\,\si{ms}$. 
    }
    \label{fig:pulse}
\end{figure*}

\textbf{Derivation of the mean-field model.}
We describe the dilute vapour of Rydberg atoms as an ensemble of $N$ two-level systems, with ground state $|g\rangle$ and Rydberg state $|e\rangle$. 
This two-level reduction follows from adiabatic elimination of the intermediate EIT state, valid at the comparatively low Rydberg densities of our experiment; at much higher densities a full three-level treatment becomes necessary~\cite{Wadenpfuhl2023Synchronization}.
The dynamics is described by a Lindblad quantum master equation for the reduced state of the system $\rho$ ($\hbar=1$):
\begin{equation}
    \partial_t \rho=-i[H,\rho]+\sum_{j=1}^N \big(\mathcal{D}_{L_{j,\mathrm{s}}}[\rho]+\mathcal{D}_{L_{j,\mathrm{d}}}[\rho]\big),
\end{equation}
where we have defined the dissipator $\mathcal{D}_{L}[\rho]=L\rho L^\dagger-\{L^\dagger L,\rho\}/2$, which models the processes of spontaneous emission and dephasing through the jump operators $L_{j,\mathrm{s}}=\sqrt{\gamma}\dyad{g}{e}_j$ and $L_{j,\mathrm{d}}=\sqrt{\gamma_\mathrm{d}}\dyad{e}{e}_j$, respectively. The system Hamiltonian reads
\begin{equation}
    H= -\Delta\sum_{j=1}^N \dyad{e}{e}_j+\frac{\Omega}{2}\sum_{j=1}^N(\dyad{e}{g}_j+\dyad{g}{e}_j)+H_\mathrm{int},
\end{equation}
where we model the interaction between atoms in the Rydberg state with an all-to-all energy shift: $ \frac{V}{N-1}\sum_{j<k}\dyad{e}{e}_j \dyad{e}{e}_k$.
In the thermodynamic limit, we consider the dynamics of the average Rydberg population, $n(t)=\frac{1}{N}\sum_{j=1}^N\ev*{\dyad{e}_j}$, and average optical coherence $q(t)=\frac{1}{N}\sum_{j=1}^N\ev*{\dyad{e}{g}_j}$, for which we obtain the following equations of motion using the mean-field approximation:
\begin{equation}
    \begin{split}
        \dot{n}&= \Omega\text{Im}[q]-\gamma n ,\\
        \dot{q}&=-i(\Delta-nV) q-\Gamma q-i\Omega\big(n-\frac{1}{2}\big),
    \end{split}
\end{equation}
where we have omitted the time labels, denoted the derivative with respect to time with a dot, and defined $\Gamma=(\gamma+\gamma_\mathrm{d})/2$. Assuming $\gamma_\mathrm{d}\gg \gamma,\Omega$, the relaxation dynamics of the average optical coherence is much faster than that of the average Rydberg population. Then, we adiabatically eliminate the coherences by setting $\dot{q}=0$ and replacing:
\begin{equation}
q=-\frac{i\Omega\big(n-\frac{1}{2}\big)}{i(\Delta-nV)+\Gamma}
\end{equation}
in the dynamics of the average Rydberg population.
This yields the mean-field equation presented in the main text, which is used for the numerical analysis presented there.

Order-of-magnitude estimates place the experiment in the collective regime that the all-to-all model is meant to capture. At \SI{47}{\celsius} the caesium density $n\approx\SI{3.61e11}{\per\cubic\centi\meter}$~\cite{vsibalic2017arc} corresponds to a mean interatomic spacing $a\approx\SI{0.8}{\micro\meter}$, while the blockade radius of the $48D_{5/2}$ pair state is $R_\mathrm{b}\approx\SIrange{2}{3}{\micro\meter}$~\cite{vsibalic2017arc}. Since $R_\mathrm{b}\gtrsim a$, of order ten to several tens of atoms lie within a single blockade volume and respond collectively rather than independently, which is precisely the regime in which a mean-field all-to-all treatment provides a faithful minimal description of the bistable transition. Consistently, the value $V/\gamma=100$ used in the model reflects a van der Waals interaction at the mean spacing ($\sim10^2$--$10^3\,$MHz) that far exceeds the effective linewidth ($\sim\,$MHz). The model fixes these dimensionless ratios so as to reproduce the macroscopic bistability rather than to resolve the distance-dependent interaction potential.

Doppler averaging and velocity-dependent interactions are omitted; the model reproduces the qualitative dependence of prediction error on detuning without them.
More microscopic extensions, such as Doppler averaging and the full three-level EIT treatment, represent natural directions for future theoretical work.

\textbf{Theoretical analysis of the relaxation times.}
Relaxation times are analysed by linearizing Eq.~(\ref{eq:n-dot-eliminated}) around each of its stationary solutions, i.e., introducing $n(t)=n_\mathrm{ss}+\delta n(t)$ and keeping only linear terms in $\delta n(t)$. This results in a linear equation for the fluctuation dynamics, $\delta\dot{n}= -\tau_\mathrm{relax}^{-1} \delta n$, where $\tau_\mathrm{relax} = (\big| \pdv{F(n)}{n}\big|_{n_\mathrm{ss}})^{-1}$ defines the characteristic relaxation time.

\textbf{Experimental analysis of the relaxation times.} In order to obtain the relaxation time of the Rydberg system $\tau_\mathrm{relax}$, we measure the transmission of the probe laser as the Rabi frequency of the coupling laser $\Omega_c$ is amplitude modulated. The amplitude modulation corresponds to a square wave with frequency $\SI{1}{\kilo\hertz}$, duty cycle 50\%, and modification depth 15\%. In Fig.~\ref{fig:pulse}(a), the detuning is swept linearly in time from $\SIrange{6}{22}{\mega\hertz}$. When the detuning crosses the bistable region, the response is enhanced, while the relaxation time increases. This is analysed in more detail in Figs.~\ref{fig:pulse}(b)-(c), where the detuning $\Delta_c$ is fixed outside or inside the bistable region, respectively, by an ultra-stable cavity and the Pound-Drever-Hall (PDH) method \cite{Black2001PDH,PDH2018sujuan,Drever1983PDH}. This allows us to obtain the relaxation time by fitting the function $A \exp(-t/\tau) + B$ to every period of the transmission signal, obtaining the values $\tau_1=0.139\,\si{ms}$, $\tau_2=0.148\,\si{ms}$ and $\tau_3=0.125\,\si{ms}$ for the data shown in panel (c). Notice that in panel (b) the relaxation times are orders of magnitude shorter.\\

\subsection{Origin of the learning performance enhancement}
\label{sec:origin-performance}
\begin{figure}[h]
    \centering
    \includegraphics[width=\linewidth]{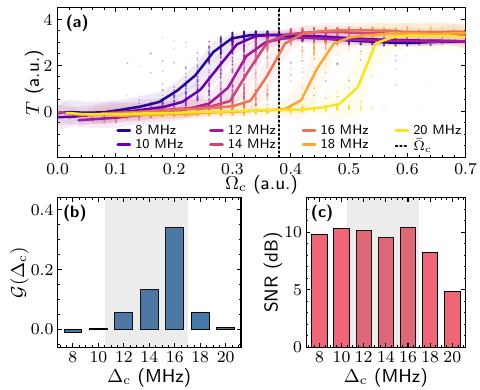}
    \caption{\textbf{Steady-state transfer curves, effective gain, and signal-to-noise ratio.}
    \textbf{(a)}~Binned transmission $\langle T \rangle$ as a function of the input drive $\Omega_\mathrm{c}$ in arbitrary units for coupling detunings $\Delta_\mathrm{c} = 8$--$20\,$MHz. Scattered points show individual measurements; solid lines are the bin means $\langle T \rangle_b$ and shaded bands the $\pm 1\sigma$ within-bin standard deviation. The vertical dashed line marks the median operating point $\bar\Omega_\mathrm{c}$ at which the effective gain in panel (b) is evaluated.
    \textbf{(b)}~Effective gain $\mathcal{G}(\Delta_\mathrm{c})$ from Eq.~\eqref{eq:effective_gain}. The collective susceptibility is sharply enhanced inside the bistable region (gray shaded interval), with a peak of $0.34$ at $\Delta_\mathrm{c}=16\,$MHz approximately thirty times larger than the values at $\Delta_\mathrm{c}=8$ and $20\,$MHz.
    \textbf{(c)}~Signal-to-noise ratio of the transfer curves from Eq.~\eqref{eq:snr}. The SNR remains high and approximately flat across $\Delta_\mathrm{c}=8$--$16\,$MHz, demonstrating that the V-shape of the prediction error is set by the gain peak in (b) rather than by a noise minimum.
    }
    \label{fig:M4_transfer_curves}
\end{figure}

To identify the physical origin of the V-shaped detuning dependence of the prediction error observed in Fig.~\ref{fig:prediction}(j) and Fig.~\ref{fig:benchmarks}, we dissect the response of the Rydberg vapour into three properties: (i) the local sensitivity of the transmission to the input, (ii) the noise level of the response, and (iii) the structure of the temporal information encoded in the reservoir. The analysis is performed on the dataset of the NARMA task, in which the coupling Rabi frequency is amplitude modulated by a discrete-step sequence drawn from a Gaussian distribution, which allows the reconstruction of the system response and the computation of the information processing capacity (IPC).

\textbf{Effective gain and signal-to-noise ratio.}
Based on the data of the NARMA task, Figure ~\ref{fig:M4_transfer_curves}(a) shows the experimentally measured transfer curve $\langle T\rangle(\Omega_\mathrm{c})$, i.e., the transmission as a function of the coupling Rabi frequency, for detunings $\Delta_\mathrm{c}=8$--$20\,$MHz.
To construct these curves, the range of $\Omega_\mathrm{c}$ values is divided into equally spaced bins; within each bin $b$, all individual transmission measurements (scattered points) are pooled to yield a bin mean $\langle T\rangle_b$ (solid lines) and a within-bin standard deviation (shaded bands).
Each curve has a sigmoidal shape, with the inflection point shifting to larger $\Omega_\mathrm{c}$ as $\Delta_\mathrm{c}$ increases.
For detunings inside the bistable region ($\Delta_\mathrm{c}=12$--$18\,$MHz), the inflection point falls close to the median operating point $\bar\Omega_\mathrm{c}$ (vertical dashed line), so the system operates near the steepest part of its transfer function.
Outside the bistable region, $\bar\Omega_\mathrm{c}$ falls on a flatter portion of the sigmoid, where the local slope is considerably smaller.

Based on these data, we evaluate the effective gain,
\begin{equation}
    \mathcal{G}(\Delta_\mathrm{c}) = \left. \frac{\partial \langle T \rangle}{\partial \Omega_\mathrm{c}} \right|_{\Omega_\mathrm{c} = \bar\Omega_\mathrm{c}},
    \label{eq:effective_gain}
\end{equation}
defined as the local slope of the binned transfer curve evaluated at the median operating point $\bar\Omega_\mathrm{c}=\mathrm{median}(\Omega_\mathrm{c}^{(i)})$ (vertical dashed line in Fig.~\ref{fig:M4_transfer_curves}(a)). As shown in Fig.~\ref{fig:M4_transfer_curves}(b), $\mathcal{G}(\Delta_\mathrm{c})$ peaks sharply inside the bistable region, reaching $0.34$ at $\Delta_\mathrm{c}=16\,$MHz. This enhancement is a direct signature of the collective amplification near the bistable phase transition: small variations of the input are translated into large, well-separated reservoir states that can be decoded by a linear model.

The second observable is the signal-to-noise ratio of the transfer curve,
\begin{equation}
    \mathrm{SNR}(\Delta_\mathrm{c}) = 10 \log_{10} \left( \frac{\sigma^2_\mathrm{between}}{\sigma^2_\mathrm{within}} \right),
    \label{eq:snr}
\end{equation}
where
\begin{align}
    \sigma^2_\mathrm{between} &= \frac{1}{B}\sum_{b=1}^B \bigl(\langle T\rangle_b - \bar{T}\bigr)^2, \quad \bar{T} = \frac{1}{B}\sum_{b=1}^B \langle T\rangle_b, \label{eq:sigma_between}\\
    \sigma^2_\mathrm{within} &= \frac{1}{B}\sum_{b=1}^B \mathrm{Var}(T_b), \label{eq:sigma_within}
\end{align}
with $B$ the total number of bins. Here $\sigma^2_\mathrm{between}$ is the variance of the bin means around their grand mean $\bar{T}$, quantifying how well the system separates distinct input levels in its output; $\sigma^2_\mathrm{within}$ is the average of the within-bin variances, capturing transmission fluctuations at a fixed input level due to optical and electronic noise. As shown in Fig.~\ref{fig:M4_transfer_curves}(c), the SNR is high and approximately flat ($9.5$--$10.5\,$dB) across $\Delta_\mathrm{c} = 8$--$16\,$MHz and drops only at the largest detunings. The noise floor is therefore essentially independent of the operating point inside the bistable window, and the V-shaped NRMSE profile cannot be ascribed to a noise minimum near the phase transition. The collective gain $\mathcal{G}$ is the dominant control parameter.

\textbf{Information processing capacity.}
The transfer curves only describe the instantaneous response. The prediction task additionally requires the reservoir to carry temporal information about the input. We quantify this by the information processing capacity~(IPC)~\cite{Dambre2012}, resolved by polynomial degree $d$ and time lag $k$. The target basis consists of Hermite polynomials evaluated on the Gaussian-distributed input sequence, which are orthogonalised by QR decomposition to prevent double-counting of capacity across different $(d,k)$ channels.

\begin{figure}[t]
    \centering
    \includegraphics[width=\linewidth]{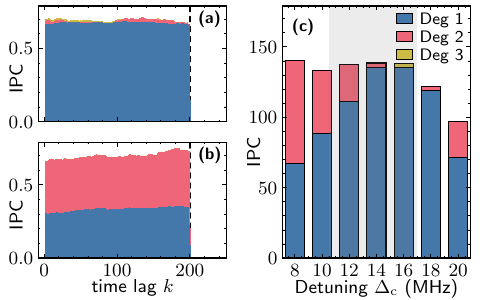}
    \caption{\textbf{Information processing capacity (IPC) of the Rydberg reservoir.}
    \textbf{(a)}~Lag-resolved IPC inside the bistable region ($\Delta_\mathrm{c}=14\,$MHz), stacked by polynomial degree (blue: linear $d=1$, red: quadratic $d=2$, yellow: cubic $d=3$). The capacity is dominated by the linear channel and remains essentially constant for all lags $k<200$, before dropping abruptly to zero at $k=200$ (vertical dashed line), the size $M$ of the sliding readout window.
    \textbf{(b)}~Lag-resolved IPC outside the bistable region ($\Delta_\mathrm{c}=8\,$MHz). About half of the capacity now resides in the quadratic channel, indicating that the input is encoded through the nonlinear tail of the EIT hysteresis loop -- a representation the linear readout cannot invert.
    \textbf{(c)}~Total IPC summed over lags $k=1$--$300$ as a function of detuning, resolved by polynomial degree. The quadratic contribution dominates at the smallest detuning, collapses to nearly zero inside the bistable region (gray shaded interval), and reappears at $\Delta_\mathrm{c}=20\,$MHz, mirroring the V-shaped NRMSE profile of the linear readout.
    }
    \label{fig:M2_ipc_combined}
\end{figure}

\begin{figure}[h]
    \centering
    \includegraphics[width=\linewidth]{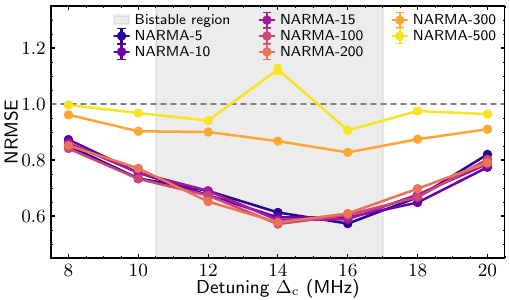}
    \caption{\textbf{NARMA performance and the readout-defined memory horizon.}
    Normalised root mean square error (NRMSE) on the NARMA-$n$ benchmark for $n=5,10,15,100,200,300,500$ as a function of coupling detuning $\Delta_\mathrm{c}$. For all tasks with memory depth $n\le 200$, the error reproduces the V-shape observed in Fig.~\ref{fig:prediction}(j), with the minimum located inside the bistable region (gray shaded interval). For $n\ge 300$, which exceeds the sliding readout window $M=200$, the NRMSE collapses to the trivial baseline NRMSE$\,\approx 0.9$ at every detuning. The transition at $n=M$ identifies the readout window, not the physical relaxation of the Rydberg vapour, as the operative memory horizon.
    }
    \label{fig:M1_narma_scans}
\end{figure}

Figures~\ref{fig:M2_ipc_combined}(a,b) show the lag-resolved IPC inside and outside the bistable region. At $\Delta_\mathrm{c}=14\,$MHz [panel~(a)] the capacity is almost entirely linear ($d=1$, blue) across all lags. At $\Delta_\mathrm{c}=8\,$MHz [panel~(b)] approximately half of the capacity is carried by the quadratic channel ($d=2$, red), with a correspondingly reduced linear contribution. The detuning dependence of the integrated capacity (summed over lags $k=1$--$300$) is summarised in Fig.~\ref{fig:M2_ipc_combined}(c): the quadratic contribution is largest at small detunings, collapses to a value compatible with zero inside the bistable region, and re-emerges at $\Delta_\mathrm{c}=20\,$MHz. Higher-order ($d\geq 3$) contributions are negligible throughout.

This basis structure explains why the linear readout layer benefits from operation inside the bistable region. Outside, the operating point $\bar \Omega_\mathrm{c}$ sits on the nonlinear tail of the EIT hysteresis loop (see Fig.~\ref{fig:M1_narma_scans}), which encodes the input through a quadratic transfer function that the linear readout cannot invert. Inside the bistable region, the same input is mapped onto an almost purely linear basis to which the linear prediction is naturally matched. We emphasise that this does not imply linear physics: the nonlinearity provided by the collective susceptibility (large $\mathcal{G}$) acts as a strong prefactor on top of the linear-dominated representation, supplying the amplification without distorting the basis.

\textbf{Memory horizon.}
Both panels (a) and (b) of Fig.~\ref{fig:M2_ipc_combined} also reveal that the total capacity stays approximately constant for lags $k<200$ and drops abruptly to zero at $k=200$. This boundary coincides exactly with the size $M=200$ of the sliding readout window. The memory available to the linear regressor is therefore controlled by the readout architecture rather than by the intrinsic relaxation of the vapour. Two further observations corroborate this picture. In particular, Fig.~\ref{fig:M1_narma_scans} shows the NRMSE on the NARMA-$n$ benchmark for $n=5,10,15,100,200,300,500$. For tasks with memory depth $n\le 200$, the error reproduces the V-shape of Fig.~\ref{fig:prediction}(j) with a clear minimum inside the bistable region. For $n\ge 300$ the error collapses to the trivial baseline NRMSE$\,\approx 0.9$ at every detuning. The sharp transition at $n=M=200$ identifies the readout window, not the relaxation of the vapour, as the operative memory horizon: tasks demanding deeper history cannot be solved at any detuning, while shorter-memory tasks fully exploit the gain peak.

\textbf{Data availability.} Data supporting this study are openly available on~\cite{Zenodo}.

\textbf{Code availability.}
The code used to generate the results in this study is openly available on Zenodo~\cite{Zenodo}. It is licensed under the GPLv3 License.

%
    
\section{Acknowledgments}

We acknowledge the National Key R\&D Program of China (Grant no. 2022YFA1404003), the National Natural Science Foundation of China (Grants T2495252, 12104279, 123B2062, 12574318), Innovation Program for Quantum Science and Technology (Grant No. 2021ZD0302100), the Fund for Shanxi ‘1331 Project’ Key Subjects Construction, Bairen Project of Shanxi Province, supported by111 project (Grant No. D18001), PCSIRT (No. IRT\_17R70). AC acknowledges support from the Deutsche Forschungsgemeinschaft (DFG, German Research Foundation) through the Walter Benjamin programme, Grant No. 519847240 and from both the Spanish Ministerio de Ciencia, Innovación y Universidades and  Universitat de les Illes Balears through the Beatriz
Galindo programme (BG24/00134). Further support was received from the DFG through the Research Units FOR 5413/1, Grant No. 465199066, and FOR 5522/1, Grant No. 499180199. 
We also acknowledge funding through JST-DFG 2024: Japanese-German Joint Call for Proposals on “Quantum Technologies” (Japan-JST-DFG-ASPIRE 2024) under DFG Grant No. 554561799. This work is also supported by the ERC grant OPEN-2QS (Grant No. 101164443).

\section*{Author contributions statement}
Z.L. conceived the idea for the study. Z.L. and Q.R. conducted the physical experiments. Z.L. and W.X. designed the learning model protocols. C.N and A.C. simulated the theoretical model. Z.L., Q.R., C.N., A.C., W.X., Y.T., H.W., W.Y., J.X., M.J., H.Z., L.X., S.J., L.Z., and I.L. analysed the data. The manuscript was written by Z.L., Q.R., C.N., A.C., H.W., and I.L. The research were supervised by L.Z. and I.L. All authors contributed to discussions regarding the results and analysis contained in the manuscript.

\section*{Additional information}
\textbf{Competing interests}. The authors declare no competing interests.

\end{document}